\documentclass[preprint,aps,showpacs,superscriptaddress]{revtex4}
\usepackage{graphicx}
\usepackage{dcolumn}
\usepackage{amsfonts, amssymb}
\begin{document}

\title{Faddeev calculations of the $\bar{K}NN$ system with
chirally-motivated $\bar{K}N$ interaction. \\
I. Low-energy $K^- d$ scattering and antikaonic deuterium.}

\author{N.V. Shevchenko\footnote{Corresponding author:
%phone: +420 266 173 276, fax: +420 220 940 165, e-mail:
shevchenko@ujf.cas.cz}}
\affiliation{Nuclear Physics Institute, 25068 \v{R}e\v{z}, Czech Republic}

\author{J. R\'{e}vai}
\affiliation{Research Institute for Particle and
Nuclear Physics, H-1525 Budapest, P.O.B. 49, Hungary}

\date{\today}

\begin{abstract}
A chirally-motivated coupled-channel $\bar{K}N$ potential,
reproducing all low-energy experimental data on $K^- p$ scattering
and kaonic hydrogen and suitable for using in accurate few-body calculations,
was constructed. The potential was used for calculations of 
low-energy amplitudes of the elastic $K^- d$ scattering
using Faddeev-type AGS equations with coupled $\bar{K}NN$ and $\pi \Sigma N$
channels. A complex $K^- - d$ potential reproducing the three-body $K^- d$
amplitudes was constructed and used for calculation of $1s$ level shift and
width of kaonic deuterium. The predicted shift $\Delta E_{1s}^{K^- d} \sim -830$
eV and width $\Gamma_{1s}^{K^- d} \sim 1055$ eV are close to our previous
results obtained with phenomenological $\bar{K}N$ potentials.
\end{abstract}

%13.75.Jz: kaon-baryon interactions
%11.80.Gw: Multichannel scattering
%36.10.Gv: Mesonic atoms and molecules, hyperonic atoms and molecules
\maketitle

%%%%%%%%%%%%%%%%%%%%%%%%%%%%%%%%%%%%%%%%%%%%%%%%%%%%%%%%%%%%%%%%
\section{Introduction}
\label{Introduction_sect}

Interaction of antikaon with nucleon is the basis for investigation
of atomic and strong quasi-bound states in antikaonic-nucleus systems.
Available two-body experimental information on $\bar{K}N$ interaction
is insufficient for construction of a unique interaction model.
In particular, it was shown in~\cite{my_Kd,my_Kd_sdvig} that phenomenological
models of the interaction having one or two poles for the $\Lambda(1405)$
resonance reproducing all low-energy
experimental data on $K^- p$ scattering and kaonic hydrogen equally
well can be constructed. A way to obtaine some additional information
about the $\bar{K}N$ interaction is to use it as an input in an accurate
few-body calculation and then compare the theoretical predictions with
eventual experimental data.

There are several calculations devoted to the low-energy $K^- d$
scattering~\cite{Kd_TGE,Kd_TDD} or the $K^- d$ scattering length
only~\cite{Kd_Deloff,Kd_BFMS_new} based on Faddeev equations.
Low-energy $K^- d$ amplitudes, including scattering length, and effective
range were calculated in our papers~\cite{my_Kd,my_Kd_sdvig}.
In the most recent one~\cite{my_Kd_sdvig} the directly measurable
characteristics of $1s$ level of kaonic deuterium
were calculated as well. It allows the direct comparison of the theoretical
predictions with eventual experimental data on kaonic deuterium, which
hopefully will be obtained in SIDDHARTA-2 experiment~\cite{SIDDHARTA-2}.

The results were obtained by solving coupled-channel Faddeev-type AGS
equations with phenomenological $\bar{K}N$ potentials. However,
many other authors of $\bar{K}N$ interaction models use not a phenomenological,
but a chirally-motivated potential, where a $\bar{K}N$ amplitude obtained
from a chiral Lagrangian is used as a potential to determine
the position of the poles of $\Lambda(1405)$ resonance. Bethe-Salpeter or
Lippmann-Schwinger equations are used for this task. There are quite a few
such chirally-motivated potentials available, however, none of them
is suited for use in Faddeev calculations since either they have too many
coupled channels and cannot be used as it is or they are not as accurate
in reproducing experimental data as one would wish. Therefore, we decided
to construct a new chirally-motivated model of $\bar{K}N$ interaction,
which can be used in dynamically accurate three-body calculations. 

The potential reproduces the low-energy data on $K^- p$ scattering and
kaonic hydrogen with the same level of accuracy as our previously constructed
phenomenological $\bar{K}N$ potentials.
We repeated our calculation of the low-energy $K^- d$ elastic scattering
and the characteristics of kaonic deuterium using the new model of $\bar{K}N$
interaction and compared the new results
with those obtained using phenomenological $\bar{K}N$ potentials.
Since the three-body AGS equations and the rest
of the two-body input are the same in both calculations, we could isolate
the pure effect of the different types of $\bar{K}N$ interaction models.

The description of the chirally motivated potential is given in the next
section, the results on the low-energy $K^- d$ scattering are given and
discussed in Sec.~\ref{Kdscat_sect}. Sec.~\ref{KaonDeu_sect} contains
information on the evaluation of kaonic deuterium $1s$ level shift and width,
while Sec.~\ref{conclusions.sect} concludes the paper.

\section{Chirally-motivated $\bar{K}N - \pi \Sigma - \pi \Lambda$ potential}
\label{ChiralKN_sect}

There are many different chirally-motivated models of $\bar{K}N$ interaction
in the literature~\cite{Borasoy_aKp,cieply2012,oller}.
Most of them are not really suited for use in Faddeev calculations
since they have too many coupled channels. Recently, one more of these $\bar{K}N$
potentials was constructed~\cite{ihw} together with a reduced version,
which contains only three channels. Therefore, this last one, in principle, could be
used in a dynamically correct few-body calculation, however, the reduced version
does not reproduce $K^- p$ experimental data accurately enough.
%as well as its more advanced variant.

The commonly used $s$-wave chirally-motivated potentials have the
energy-dependent part (see e.g.~\cite{cieply2012})
\begin{equation}
\label{VchiralPart}
\bar{V}^{ab}(\sqrt{s} ) =
  \sqrt{ \frac{M_{a}}{2 \omega_{a} E_{a} }} \,
  \frac{C^{ab}(\sqrt{s})}{(2 \pi)^3 f_{a} f_{b}} \,
  \sqrt{ \frac{M_{b}}{2 \omega_{b} E_{b} }}
\end{equation}
and are written in particle basis. We took into account all open particle
channels: $a, b = K^- p, \bar{K}^0n, \pi^+ \Sigma^-, \pi^0 \Sigma^0, \pi^- \Sigma^+$
and $\pi^0 \Lambda$. Baryon mass $M_{a}$, baryon energy $E_{a}$ and meson energy
$w_{a}$ of the channel $a$ enter the factors, which ensure
proper normalization of the amplitude. The non-relativistic form of the leading
order Weinberg-Tomozawa interaction
\begin{equation}
C^{ab}(\sqrt{s}) = 
 - C^{WT} \, (2\sqrt{s} - M_{a} - M_{b})
\end{equation}
was used with SU(3) Clebsh-Gordan coefficients $C^{WT}_I$.

Our chirally motivated potential $V^{Chiral}_{\bar{K}N}$ is separable,
it contains also form-factors and is written in isospin basis:
\begin{equation}
\label{VchiralIso}
V_{I I'}^{\alpha \beta}(k^{\alpha},k'^{\beta};\sqrt{s} ) =
  g_I^{\alpha} (k^{\alpha}) \, \bar{V}_{I I'}^{\alpha \beta}(\sqrt{s}) \,
  g_{I'}^{\beta} (k'^{\beta}) \,,
\end{equation}
where $V_{I I'}^{\alpha \beta}(\sqrt{s})$ is the energy-dependent part of the
potential in isospin basis, obtained from Eq.(\ref{VchiralPart}). 
The $k^{\alpha}, k'^{\alpha}$ and $\sqrt{s}$ stand for the initial,
final relative momenta and the total energy, respectively. We used
physical masses in the calculations, therefore the two-body isospin
$I = 0$ or $1$ is not conserved. Yamaguchi form-factors
\begin{equation}
g_I^{\alpha} (k^{\alpha}) = 
 \frac{(\beta_I^{\alpha})^2}{(k^{\alpha})^2 + (\beta_I^{\alpha})^2}
\end{equation}
were used in~Eq.(\ref{VchiralIso}). The channel indices $\alpha, \beta$
take three values denoting the $\bar{K}N$, $\pi \Sigma$ and $\pi \Lambda$
channels.

The pseudo-scalar meson decay constants $f_{\pi}$, $f_K$ and the range parameters
$\beta_I^{\alpha}$, depending on the two-body isospin, are free
parameters, which were found by fitting the potential to the experimental
data. In the same way as the phenomenological ones, the potential~(\ref{VchiralIso})
reproduces elastic and inelastic $K^- p$ cross-sections, threshold branching
ratios $\gamma$, $R_c$, $R_n$  and characteristics of $1s$ level of kaonic hydrogen.
The parameters of the potential are shown in Table~\ref{params.tab}.
%----------------------------------------------------------------
\begin{center}
\begin{table}
\caption{Parameters of the chirally-motivated
$V^{Chiral}_{\bar{K}N - \pi \Sigma - \pi \Lambda}$
potential: the pseudo-scalar meson decay constants $f_{\pi}$, $f_K$
(MeV) and the range parameters $\beta_I^{\alpha}$ (fm${}^{-1}$).}
\label{params.tab}
\begin{center}
\begin{tabular}{ccccccc}
\hline  \noalign{\smallskip}
 $f_{\pi}$ & $f_K$  & 
   \quad $\beta^{\bar{K}N}_0$  & \quad $\beta^{\pi \Sigma}_0$  &
   \quad $\beta^{\bar{K}N}_1$  & \quad $\beta^{\pi \Sigma}_1$  & 
   \quad $\beta^{\pi \Lambda}_1$ \\
\noalign{\smallskip} \hline \noalign{\smallskip}
 116.20 \quad & 113.36 \quad & 4.06 \quad & 3.30 \quad 
  & 5.00 \quad & 3.86 \quad & 1.99 \\
\noalign{\smallskip} \hline
\end{tabular}
\end{center}
\end{table}
\end{center}
%----------------------------------------------------------------

All physical observables to be compared with experimental data were obtained
from solution of the Lippmann-Schwinger equation with the potential
$V^{Chiral}_{\bar{K}N}$~(\ref{VchiralIso}) and Coulomb interaction since, as
previously, we wanted to calculate characteristics of kaonic hydrogen directly,
without intermediate reference to $K^- p$ scattering length. 
We used non-relativistic kinematics while the potential was constructed.
Among all authors of $\bar{K}N$ potentials only we and
Cieply, Smejkal~\cite{cieply2007,cieply2012}
take Coulomb interaction into account directly when 
calculating the $1s$ level shift and width of kaonic hydrogen.
All other calculations of the same quantity get it from the $K^- p$
scattering length through the approximate ``corrected Deser''
formula~\cite{corDeser}. However, as it was shown
in~\cite{ourPRC_isobreak,cieply2007} the approximate formula gives
$10 \%$ error, therefore the direct calculation of the $1s$ level shift
and width of kaonic hydrogen is desirable.
%----------------------------------------------------------------
\begin{center}
\begin{table}
\caption{Physical characteristics of the chirally motivated
$V^{Chiral}_{\bar{K}N - \pi \Sigma-\pi \Lambda}$ potential: 
$1s$ level shift $\Delta E_{1s}^{K^- p}$ (eV) and
width $\Gamma_{1s}^{K^- p}$ (eV) of kaonic hydrogen,
threshold branching ratios $\gamma$, $R_c$ and $R_n$ together
with experimental data. The experimental data on kaonic hydrogen are
those obtained by SIDDHARTA collaboration.}
\label{phys_char.tab}
\begin{tabular}{ccc}
\hline \noalign{\smallskip}
 & $V^{Chiral}_{\bar{K}N - \pi \Sigma-\pi \Lambda}$ & Experiment  \\
\noalign{\smallskip} \hline \noalign{\smallskip}
$\Delta E_{1s}^{K^- p}$  & $-313$  & $-283 \pm 36 \pm 6$~\cite{SIDDHARTA} \\
$\Gamma_{1s}^{K^- p}$    & $561$ & $541 \pm 89 \pm 22$~\cite{SIDDHARTA} \\
$\gamma$ & 2.35 & $2.36 \pm 0.04$~\cite{gammaKp1,gammaKp2} \\
$R_c$ & 0.663 & $0.664 \pm 0.011$~\cite{gammaKp1,gammaKp2} \\
$R_n$ & 0.191 & $0.189 \pm 0.015$~\cite{gammaKp1,gammaKp2} \\
\noalign{\smallskip} \hline
\end{tabular}
\end{table}
\end{center}
%----------------------------------------------------------------

The observables given by the potential are shown in Table~\ref{phys_char.tab}
together with the corresponding experimental data. It is seen that
the $1s$ level shift  $\Delta E_{1s}^{K^- p}$ and width $\Gamma_{1s}^{K^- p}$
of kaonic hydrogen of the $V^{Chiral}_{\bar{K}N}$ are in agreement with
the most recent experimental data of SIDDHARTA collaboration~\cite{SIDDHARTA}.
Comparing the data in Table~\ref{phys_char.tab} with those from Table 2
of~\cite{my_Kd_sdvig} we see that the chirally motivated potential
$V^{Chiral}_{\bar{K}N}$ gives $1s$ level shift  $\Delta E_{1s}^{K^- p}$ and
width $\Gamma_{1s}^{K^- p}$ of kaonic hydrogen, which is close to the results of
the one-pole $V^{1,SIDD}_{\bar{K}N-\pi \Sigma}$ and the two-pole
$V^{2,SIDD}_{\bar{K}N-\pi \Sigma}$ versions of the phenomenological potential.
The chirally motivated potential also reproduces the rather accurately measured
threshold branching ratios $\gamma$, $R_c$ and $R_n$:
\begin{equation}
\label{gamma}
\gamma =
 \frac{\Gamma(K^- p \to \pi^+ \Sigma^-)}{\Gamma(K^- p \to
    \pi^- \Sigma^+)}  \,,
\end{equation}
\begin{eqnarray}
\label{Rc}
R_c &=& \frac{\Gamma(K^- p \to \pi^+ \Sigma^-, \pi^- \Sigma^+)}{\Gamma(K^- p \to
\mbox{all inelastic channels} )} \,, \\
\label{Rn}
R_n &=& \frac{\Gamma(K^- p \to \pi^0 \Lambda)}{\Gamma(K^- p \to
\mbox{neutral states} )}  \,.
\end{eqnarray}
The medium value of the threshold branching ratio
$\gamma$ and of the $R_{\pi \Sigma}$ constructed from the $R_c$ and $R_n$
(see Eqs.(7,10) of~\cite{my_Kd_sdvig}) are reproduced by the phenomenological
potential as well, therefore, we can say that all three potentials reproduce
the experimental data equally well.

\begin{figure*}
\centering
\includegraphics[width=0.85\textwidth]{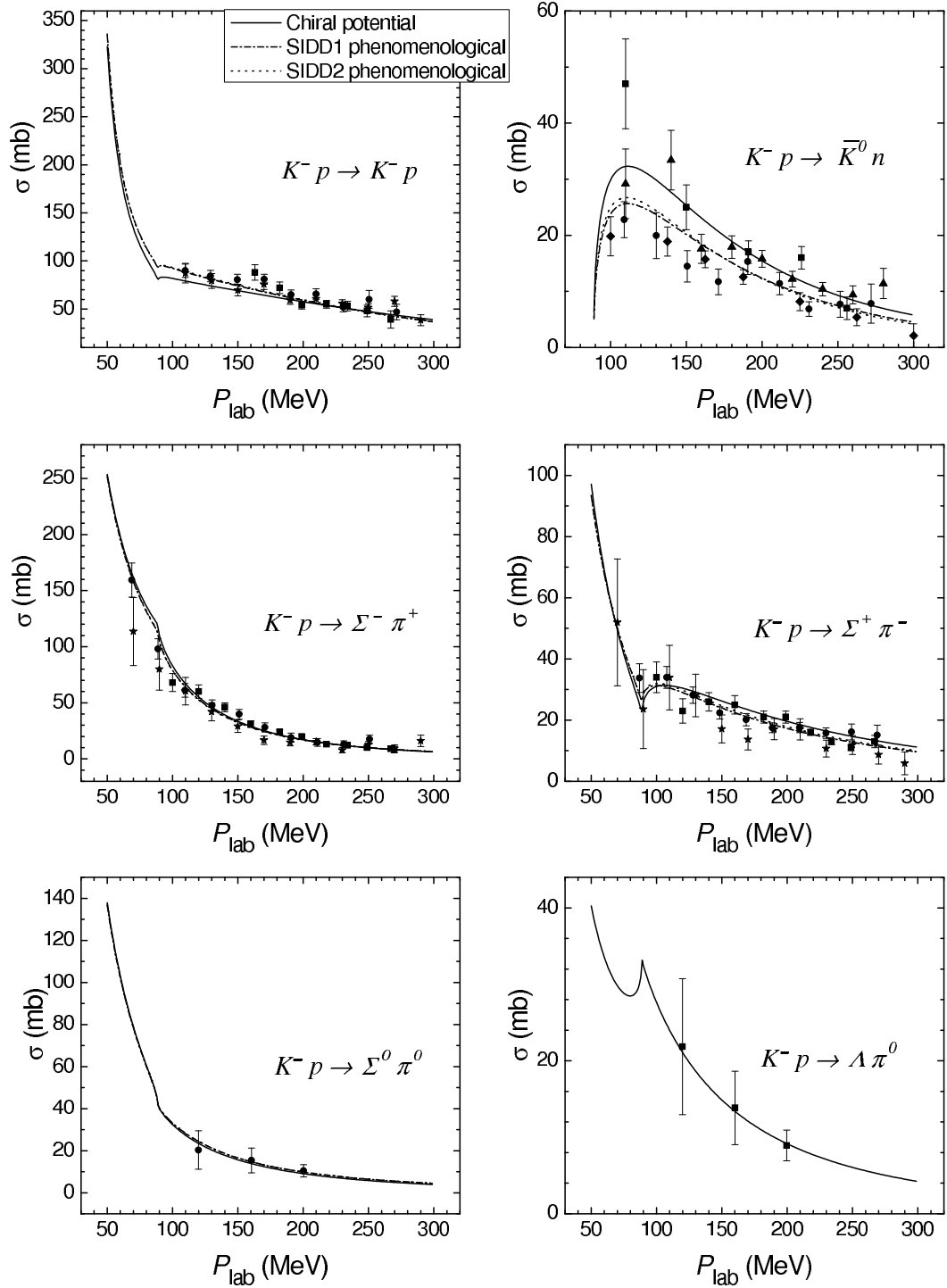}
\caption{Comparison of the elastic and inelastic $K^- p$
cross-sections for the chirally motivated potential $V^{Chiral}_{\bar{K}N}$
(solid lines) with the one-pole $V^{1,SIDD}_{\bar{K}N-\pi \Sigma}$
(dash-dotted lines) and two-pole $V^{1,SIDD}_{\bar{K}N-\pi \Sigma}$
(dotted lines) phenomenological potentials from~\cite{my_Kd_sdvig}.
The experimental data are taken from~\protect\cite{Kp2exp,Kp3exp,Kp4exp,Kp5exp,Kp6exp}
(data points).
\label{Kp_cross_sect.fig}}
\end{figure*}

The same is true for the elastic and inelastic $K^- p$ cross-sections
$K^- p \to {K}^- p$, $K^- p \to \bar{K}^0 n$, $K^- p \to \pi^+ \Sigma^-$,
$K^- p \to \pi^- \Sigma^+$, and $K^- p \to \pi^0 \Sigma^0$. In order
to demonstrate, that all three potentials reproduce the cross-sections
with the same accuracy, we plotted the results of
$V^{Chiral}_{\bar{K}N}$, $V^{1,SIDD}_{\bar{K}N-\pi \Sigma}$
and $V^{2,SIDD}_{\bar{K}N-\pi \Sigma}$ model of interaction in the same figure,
see Figure~\ref{Kp_cross_sect.fig}. The experimental data in the figure are
taken from~\cite{Kp2exp,Kp3exp,Kp4exp,Kp5exp,Kp6exp}. As previously, one
set of data~\cite{Kp1exp} is neglected due to large experimental errors.
%%%%%%%%%%%%%%%%%%%%%%%%%%%%%%%%%%%%%%%%%%%%%%%%%%%%%%%%%%%%%%%%%%%%%%%%%%
%%%%%%%%%%%%%%%%%%%%%%%%%%%%%%%%%%%%%%%%%%%%%%%%%%%%%%%%%%%%%%%%%%%%%%%%%%

Unlike most of the authors of models of $\bar{K}N$ interaction
we need not know $K^- p$ scattering length $a_{K^- p}$ to calculate the
characteristics of kaonic hydrogen. However, we can calculate it directly from
the $V^{Chiral}_{\bar{K}N}$ potential, its value is
\begin{equation}
\label{aKp}
a_{K^- p} = -0.77 + i \, 0.84 \; {\rm fm} \,.
\end{equation}
The isospin zero and one $\bar{K}N$ scattering lengths
\begin{equation}
\label{aKN01}
a_{\bar{K}N,0} = -1.65 + i \, 1.26 \; {\rm fm}, \quad
a_{\bar{K}N,1} =  0.52 + i \, 0.48 \; {\rm fm}
\end{equation}
are not connected with the $a_{K^- p}$ value by a simple formula since
physical masses are used while the $V^{Chiral}_{\bar{K}N}$ is constructed
together with isospin nonconserving Coulomb interaction. Put in three-body
AGS equations, however, isospin averaged masses are used, which lead to
different values of the scattering lengths:
\begin{eqnarray}
\label{aKNaver}
a^{aver}_{K^- p} = -0.49 + i \, 0.71 \; {\rm fm} \,, &{}&\\
a^{aver}_{\bar{K}N,0} = -1.50 + i \, 0.84 \; {\rm fm}, &\quad&
a^{aver}_{\bar{K}N,1} =  0.53 + i \, 0.59 \; {\rm fm}.
\end{eqnarray}

In the same way as other chirally-motivated potentials, our
new potential has two strong poles for the $\Lambda(1405)$ resonance:
\begin{equation}
z_1 = 1417 - i \, 33 \; {\rm MeV}, \quad
z_2 = 1406 - i \, 89 \; {\rm MeV}.
\end{equation}
Both are situated on the proper Riemann sheets, corresponding to a resonance
in $\pi \Sigma$ channel and a quasi-bound state in $\bar{K}N$ channel.
They are connected to $\pi \Lambda$ channel too through isospin nonconserving
parts. The real parts of the poles are situated between the
$\bar{K}N$ and $\pi \Sigma$ thresholds as one would expect.
The two-pole structure of the $\Lambda(1405)$ resonance follows from the
energy-dependent form of the potential. To achieve the same property of
our two-pole phenomenological potential $V^{2,SIDD}_{\bar{K}N-\pi \Sigma}$
we used a more complicated form-factor in the $\pi \Sigma$ channel.
%%%%%%%%%%%%%%%%%%%%%%%%%%%%%%%%%%%%%%%%%%%%%%%%%%%%%%%%%%%%%%%%%%%%%%%%%%
\begin{figure*}
\centering
\includegraphics[width=0.35\textwidth, angle=-90]{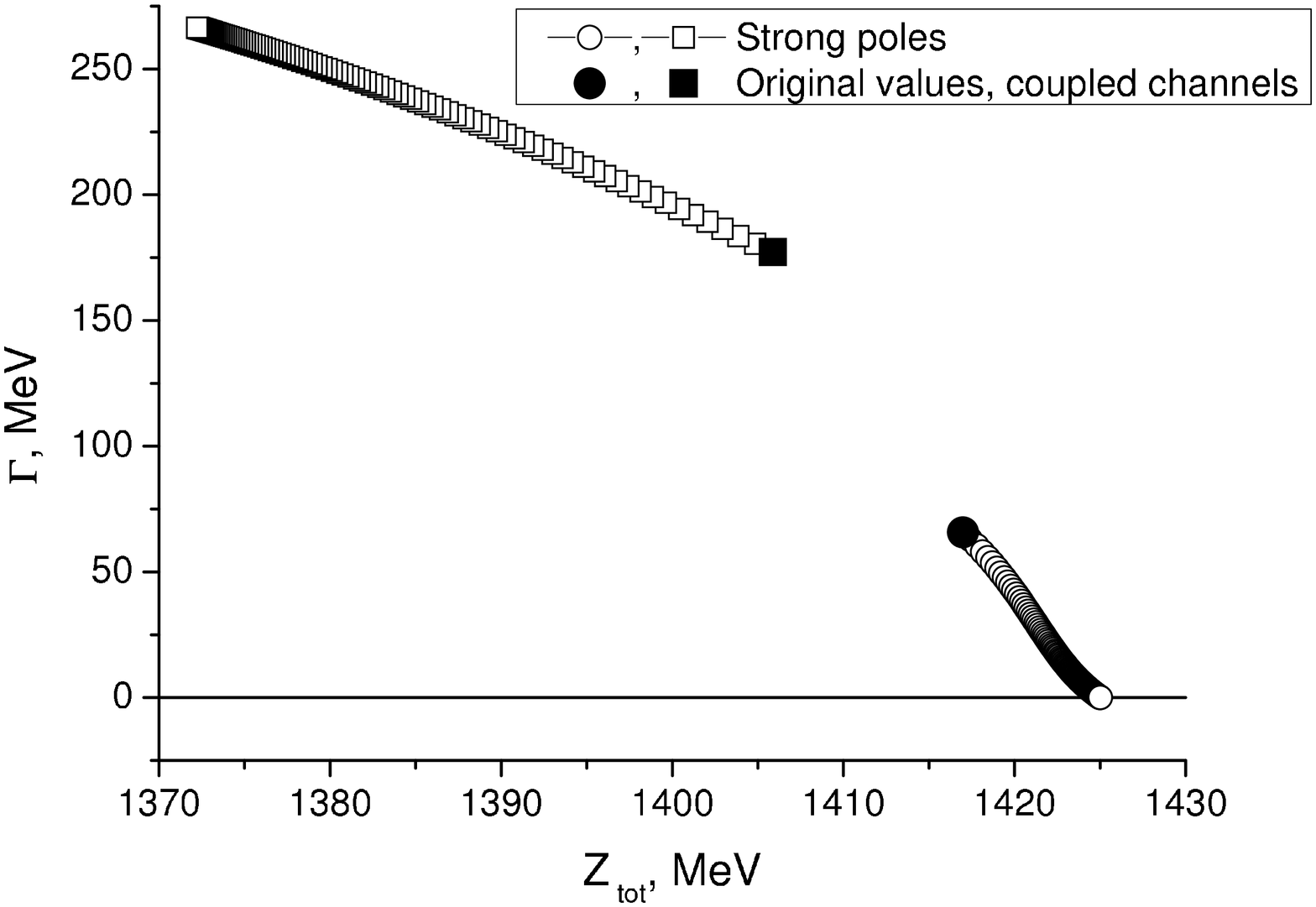}
\caption{Trajectories of the strong poles when the coupling
between the $\bar{K}N$, $\pi \Sigma$ and $\pi \Lambda$ channels is
gradually being switched off (empty symbols). The filled symbols denote
the original values with coupled channels.
\label{StrongPoles_PoleMove.fig}}
\end{figure*}
%%%%%%%%%%%%%%%%%%%%%%%%%%%%%%%%%%%%%%%%%%%%%%%%%%%%%%%%%%%%%%%%%%%%%%%%%%
%%%%%%%%%%%%%%%%%%%%%%%%%%%%%%%%%%%%%%%%%%%%%%%%%%%%%%%%%%%%%%%%%%%%%%%%%%
\begin{figure*}
\centering
\includegraphics[width=0.35\textwidth, angle=-90]{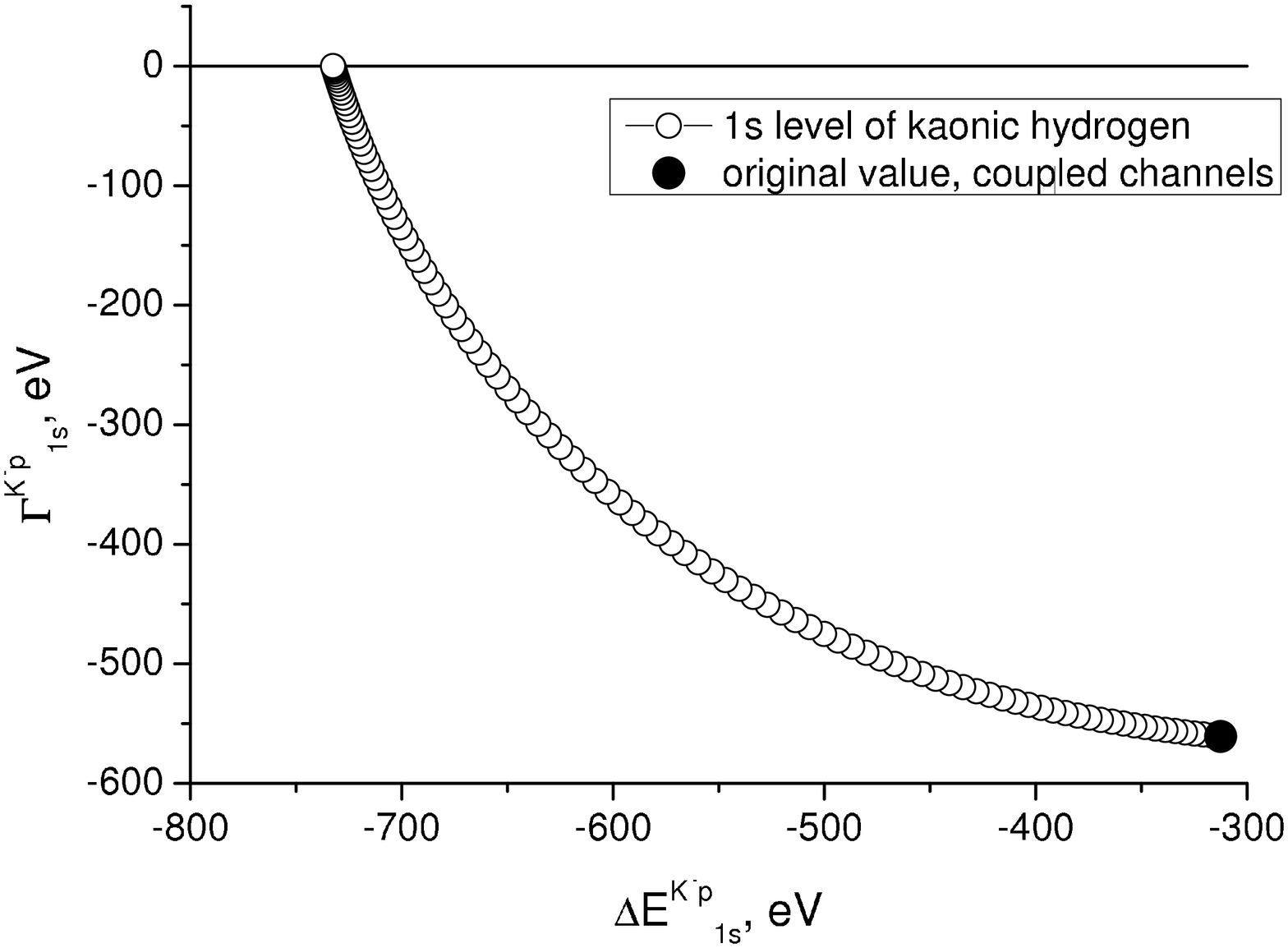}
\caption{The same as Fig.~\ref{StrongPoles_PoleMove.fig} for the
$1s$ level of kaonic hydrogen.
\label{KpAtom_PoleMove.fig}}
\end{figure*}
%%%%%%%%%%%%%%%%%%%%%%%%%%%%%%%%%%%%%%%%%%%%%%%%%%%%%%%%%%%%%%%%%%%%%%%%%%

In the same way as in~\cite{ourPRC_isobreak} we checked, where the poles move
when the nondiagonal couplings of the potential were gradually reduced to zero.
The results are demonstrated in Fig.~\ref{StrongPoles_PoleMove.fig}. It is seen,
that the strong pole $z_1$ becomes 
a real bound state with smaller than the original binding energy
when the $\bar{K}N$, $\pi \Sigma$ and $\pi \Lambda$ channels are uncoupled.
The second strong pole $z_2$ remains a
resonance pole, situated between the $\bar{K}N$ and $\pi \Sigma$ channels,
but with smaller real and larger imaginary parts. The same trajectory drawn
for the $1s$ level shift and width of kaonic hydrogen, see
Fig.~\ref{KpAtom_PoleMove.fig}, shows that the pole, corresponding to the
atomic state, also becomes a real bound state. The $1s$ level shift
is large for the decoupled system.

\begin{figure*}
\centering
\includegraphics[width=0.35\textwidth, angle=-90]{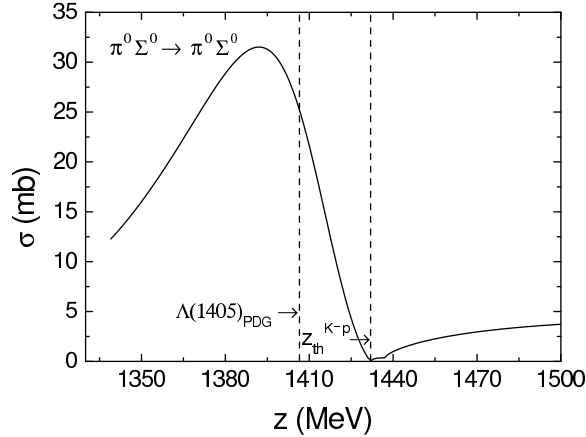}
\caption{Elastic $\pi^0 \Sigma^0$ cross-sections of the
chirally motivated potential $V^{Chiral}_{\bar{K}N}$. The PDG value
for the mass of $\Lambda(1405)$ resonance and the $K^- p$ threshold
are also shown.
\label{pi0Sig0.fig}}
\end{figure*}

Theoretically, the $\Lambda(1405)$ resonance peak could be seen in the elastic
$\pi^0 \Sigma^0$ cross-sections, however, the corresponding experimental peak can
be observed only as an FSI peak in a more complicated reaction involving
three or more particles. In this case the virtual
$\bar{K}N \to \pi \Sigma$ process can also contribute to the $\pi \Sigma$ yield
in a final state. The extent of this contribution can be reliably determined only by
considering the complete, rather complicated process (see e.g. Eq.(11) in~\cite{Janos}).
Instead, many authors of $\bar{K}N$ interaction models add
the $\bar{K}N \to \pi \Sigma$ amplitude to the $\pi \Sigma \to \pi \Sigma$
one and introduce an adjustable parameter in front of it to compare the theoretical
predictions with experimental $\pi \Sigma$ missing mass spectra. The corresponding
cross-sections are multiplied by $\pi \Sigma$ relative momentum, which is a phase
space factor coming from the FSI formalism. We did not follow that routine and
demonstrate the effect of $\Lambda(1405)$ resonance in elastic $\pi^0 \Sigma^0$
cross-sections, see Figure~\ref{pi0Sig0.fig}.
%%%%%%%%%%%%%%%%%%%%%%%%%%%%%%%%%%%%%%%%%%%%%%%%%%%%%%%%%%%%%%%%%%%%%%%%%%
%%%%%%%%%%%%%%%%%%%%%%%%%%%%%%%%%%%%%%%%%%%%%%%%%%%%%%%%%%%%%%%%%%%%%%%%%%

%-----------------------------------------------------------------------------
\section{$K^- d$ elastic scattering and $K^- d$ quasi-bound state}
\label{Kdscat_sect}

We solved Faddeev-type equations in Alt-Grassberger-Sandhas form
for the $\bar{K}NN$ system with coupled $\pi \Sigma N$ channel,
the formulas can be found in our previous paper~\cite{my_Kd}.
The equations properly describe three-body dynamics of the system.
They are written in momentum representation, isospin formalism is
used. The equations were properly antisymmetrised, which is necessary
due to two baryons in every channel. The logarithmic singularities 
in the kernels of the equations were treated by the method suggested
in~\cite{log_sing}. The three-body calculations were performed without
taking Coulomb interaction into account since the effect of it is
expected to be small.

The elastic $K^- d$ amplitudes, including the scattering length, and
effective range were calculated using the chirally-motivated
$V^{Chiral}_{\bar{K}N - \pi \Sigma - \pi \Lambda}$ potential
described in the previous section. We used averaged masses in
the potential as well as in the the whole three-body calculation 
since it was shown in~\cite{Janos} that 
the effect of physical masses is rather small.
The three-channel $\bar{K}N - \pi \Sigma - \pi \Lambda$ potential
was used in the $\bar{K}NN - \pi \Sigma N$ AGS equations in a form
of the exact optical two-channel $\bar{K}N - \pi \Sigma (- \pi \Lambda)$
potential, when the $\bar{K}N-\bar{K}N$, $\bar{K}N-\pi \Sigma$ and
$\pi \Sigma - \pi \Sigma$ elements of the three-channel $T$-matrix
are used as the two-channel $T$-matrix. The remaining
two-body potentials, needed for the three-body calculation, are also separable.
The two-term TSA-B $NN$ and the exact optical $\Sigma N (-\Lambda N)$ potentials,
which were used, are described in~\cite{my_Kd}. The $NN$ model of interaction
reproduces phase shifts of Argonne V18 potential, therefore, is repulsive
at short distances. It gives proper $NN$ scattering length, effective
range and binding energy of deuteron. The two-channel $\Sigma N -\Lambda N$
potential reproduces experimental $\Sigma N$ and $\Lambda N$ cross-sections,
the corresponding exact optical $\Sigma N (-\Lambda N)$ potential has
exactly the same elastic $\Sigma N$ amplitude as the two-channel potential.

The $K^- d$ scattering length $a_{K^- d}$ obtained with the chirally-motivated
potential is shown in Table~\ref{aKd.tab}. The new three-body result is compared
to those from~\cite{my_Kd_sdvig} with one- $V^{1,SIDD}_{\bar{K}N-\pi \Sigma}$
and two-pole $V^{2,SIDD}_{\bar{K}N-\pi \Sigma}$ versions of phenomenological
$\bar{K}N$ potential. The ``phenomenological'' results in the Table slightly differ
from the three-body values from Table~2 of~\cite{my_Kd_sdvig} since here we used
the spin-independent $\Sigma N (-\Lambda N)$ potential, while the spin-dependent
was used in the previous paper. It is seen, that the chirally motivated potential
leads to about $6 \%$ larger absolute value of the real and the imaginary part
of the scattering length. The difference is quite small, so we can conclude, that
the three different models of $\bar{K}N$ interaction, which reproduce low-energy
data on $K^- p$ scattering and kaonic hydrogen with the same level of accuracy,
give quite similar results for low-energy $K^- d$ scattering.

Since it was shown~\cite{my_Kd} that Fixed Scatterer Approximation,
also called Fixed Center Approximation, gives very large error for the
$K^- d$ system, this time we do not compare the results obtained with this
method with ours. Four $a_{K^- d}$ values obtained in other Faddeev calculations
are shown in Table~\ref{aKd.tab}. Comparing to them, we see that the result
of the very recent calculation with coupled channels~\cite{Kd_BFMS_new} gives
real part of $a_{K^- d}$, which almost coincides with our result for chirally
motivated potential. The imaginary part of the $K^- d$ scattering length
from~\cite{Kd_BFMS_new} is
slightly larger, which can follow form the fact that the model of $\bar{K}N$
interaction used there was fitted to kaonic hydrogen data not directly, but
through the $K^- p$ scattering length and the approximate formula, which 
is the least reliable just in reproducing the imaginary part of the level shift.

The one-channel result of Faddeev calculation~\cite{Kd_Deloff} lies far away
from all the others. Two effects play their role here: a one-channel dynamics
and, therefore, indirect taking $\pi \Sigma N$ channel into account and
problems with reproducing experimental data by the complex $\bar{K}N$ potential.
It was demonstrated in~\cite{my_Kd} for phenomenological models of $\bar{K}N$
interaction that simple complex potentials have quite large error, while an
exact optical $\bar{K}N$ potential gives rather accurate result for the
scattering length. The exact optical potential with $\bar{K}N$ amplitudes exactly
corresponding to those from the potential with coupled channels is good for
the chirally motivated model as well. It gives
\begin{equation}
\label{aKdOpt}
a_{K^- d}^{Chiral,Opt} = -1.57 + i \, 1.32 \; {\rm fm} \,,
\end{equation}
which is very close to the coupled-channel result from Table~\ref{aKd.tab}.

Finally, two old $a_{K^- d}$ values~\cite{Kd_TDD,Kd_TGE} significantly
underestimate the imaginary part of the $K^- d$ scattering length.
%----------------------------------------------------------------
\begin{center}
\begin{table}
\caption{Scattering lengths of $K^- d$ scattering $a_{K^- d}$ (fm) and
effective range $r^{eff}_{K^- d}$ (fm)
obtained from AGS calculations with the chirally-motivated
$V^{Chiral}_{\bar{K}N - \pi \Sigma - \pi \Lambda}$ potential (\ref{VchiralIso})
and the one-pole $V^{1,SIDD}_{\bar{K}N-\pi \Sigma}$ and two-pole 
$V^{2,SIDD}_{\bar{K}N-\pi \Sigma}$ phenomenological potentials
from~\cite{my_Kd_sdvig}. $K^- d$ scattering length values from other
Faddeev calculations are also shown. $1s$ level shift $\Delta E^{K^- d}_{1s}$
(eV) and width $\Gamma^{K^- d}_{1s}$ (eV) of kaonic deuterium, calculated using
the three potentials, are shown as well.}
\label{aKd.tab}
\begin{center}
\begin{tabular}{ccccc}
\hline \noalign{\smallskip}
 & $a_{K^- d}$ & $r^{eff}_{K^- d}$ 
  & $\Delta E^{K^- d}_{1s}$ & \quad $\Gamma^{K^- d}_{1s}$\\
\hline \noalign{\smallskip}
 AGS with $V_{\bar{K}N}^{\rm Chiral}$, this work & $-1.59 +  i \, 1.32$ 
  \qquad & $0.50 -  i \, 1.17$ & -828 & 1055 \\
 AGS with $V^{1,SIDD}_{\bar{K}N}$, \cite{my_Kd_sdvig}  & $-1.49 + i \, 1.24$ 
   \qquad & $0.69 -  i \, 1.31$ & -785 & 1018 \\
 AGS with $V^{2,SIDD}_{\bar{K}N}$, \cite{my_Kd_sdvig}  & $-1.51 + i \, 1.25$ 
   \qquad & $0.69 -  i \, 1.34$ & -797 & 1025 \\
\noalign{\smallskip} \hline \noalign{\smallskip}
 MFST, \cite{Kd_BFMS_new} & $-1.58 + i \, 1.37$ & & & \\
 Deloff, \cite{Kd_Deloff} & $-0.85 + i \, 1.05$ & & &  \\
 TDD, \cite{Kd_TDD} & $-1.34 + i \, 1.04$ & & &  \\
 TGE, \cite{Kd_TGE} & $-1.47 + i \, 1.08$ & & &  \\
\noalign{\smallskip} \hline
\end{tabular}
\end{center}
\end{table}
\end{center}
%----------------------------------------------------------------

\begin{figure*}
\centering
\includegraphics[width=0.35\textwidth, angle=-90]{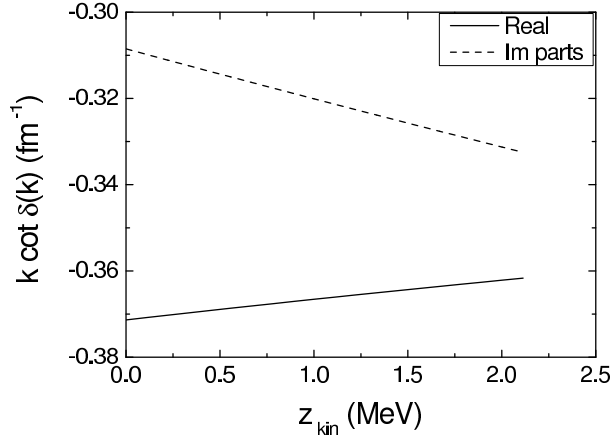}
\caption{Real (solid line) and imaginary (dashed line) parts of
the elastic near-threshold $K^- d$ amplitudes presented in a form of
$k \cot \delta(k)$ function. The results were obtained from the
coupled-channel three-body AGS equations using the 
chirally motivated potential $V^{Chiral}_{\bar{K}N}$.
\label{kCtgDelta.fig}}
\end{figure*}

We also calculated effective range $r^{eff}_{K^- d}$ of $K^- d$ scattering, 
the results can be seen in Table~\ref{aKd.tab}. The real part of
$r^{eff}_{K^- d}$ of the chirally
motivated potential is much smaller than those of our phenomenological
potentials. The imaginary part is smaller by the absolute value.
The near-threshold elastic amplitudes of $K^- d$
scattering are needed for construction of a complex two-body $K^- - d$
potential and further calculation of $1s$ level of kaonic deuterium.
They are presented in $k\, {\rm cot}\, \delta(k)$ form in Figure~\ref{kCtgDelta.fig}.
%%%%%%%%%%%%%%%%%%%%%%%%%%%%%%%%%%%%%%%%%%%%%%%%%%%%%%%%%%%%%%%%%%%%%%%%%%
%%%%%%%%%%%%%%%%%%%%%%%%%%%%%%%%%%%%%%%%%%%%%%%%%%%%%%%%%%%%%%%%%%%%%%%%%%

The relative values of $|{\rm Re}\, a_{K^- d}|$ and 
$|{\rm Im}\, a_{K^- d}|$ obtained with all three our potentials together with
their signs might lead to the conclusion that a bound or a quasi-bound
state could exist in the $K^- d$ system. However, this conclusion is not
true. We solved an eigenvalue problem using the same AGS equations and
found that such a state does not exist. Details of the calculations can be
found in our next paper, devoted to the quasi-bound state in $K^- pp$ system.
The reason why the estimation
is not reliable is that it is based on the effective range expansion of
a scattering amplitude. The expansion allows to derive simple relations
between a scattering length and a bound state of a system, however, its
validity is limited to the vicinity of the corresponding threshold. 
Since the $K^- d$ state is expected to have rather large width, it is
definitely out of such a region. The only $\bar{K}N$ potential, which
gives a quasi-bound state in $K^- d$ system, is one of our older phenomenological
potentials~\cite{my_Kd}, which does not reproduce SIDDHARTA, but KEK data
on kaonic hydrogen~\cite{KEK} only.

\section{Characteristics of kaonic deuterium}
\label{KaonDeu_sect}

Our aim was to calculate a physical quantity, which characterizes
low-energy properties of $K^- d$ system and can be compared to
experimental data directly. The scattering length is not of this type,
while the $1s$ level shift and width of kaonic deuterium can be measured.
Therefore, we calculated these atomic observables, which
correspond to the results of our three-body calculations of
low-energy $K^- d$ scattering.

Since Faddeev calculation with Coulomb plus strong interaction is too hard,
a two-body calculation with a complex $K^- - d$ potential was performed
instead. The potential is a separable one with two terms
\begin{equation}
\label{VKd}
V_{K^- d}(\vec{k},\vec{k}') = \lambda_{1,K^- d} \, g_1(\vec{k}) g_1(\vec{k}')
+ \lambda_{2,K^- d} \, g_2(\vec{k}) g_2(\vec{k}')
\end{equation}
and Yamaguchi form-factors
\begin{equation}
\label{gKd}
g_i(k) = \frac{1}{\beta_{i,K^- d}^2 + k^2}, \qquad i=1,2.
\end{equation}
The parameters of the potential
\begin{eqnarray}
 \beta_{1,K^-d} = 1.5 \; {\rm fm^{-1},} &{}& 
  \lambda_{1,K^-d} = -0.0628 - i \, 0.4974 \; {\rm fm^{-2}}\\
 \beta_{2,K^-d} = 1.1, \; {\rm fm^{-1},} &{}& 
  \lambda_{2,K^-d} = -0.1123 + i \, 0.1556 \; {\rm fm^{-2}}
\end{eqnarray}
were fixed by fitting the three-body $K^- d$ amplitudes calculated
using the AGS equations, described in the previous section. 
Obviously, the potential~(\ref{VKd}) reproduces the scattering
length $a_{K^- d}$ and effective range $r^{eff}_{K^- d}$ from
Table~\ref{aKd.tab}.

The Lippmann-Schwinger equation with the complex $K^- - d$
and Coulomb potentials was then solved and the $1s$ level energy
was obtained. More details about the calculation can be found in~\cite{my_Kd_sdvig}.
The shift $\Delta E^{K^- d}_{1s}$ and width $\Gamma^{K^- d}_{1s}$
of kaonic deuterium, corresponding to the chirally motivated model of
$\bar{K}N - \pi \Sigma - \pi \Lambda$ interaction are shown in
Table~\ref{aKd.tab}. We also show the characteristics of the atom,
obtained with our phenomenological potentials
$V^{1,SIDD}_{\bar{K}N-\pi \Sigma}$ and
$V^{2,SIDD}_{\bar{K}N-\pi \Sigma}$~\cite{my_Kd_sdvig}.

The ``chirally motivated'' absolute values of the level shift
$\Delta E_{1s}^{K^- d}$ and the width $\Gamma_{1s}^{K^- d}$ are both larger
than those obtained in~\cite{my_Kd_sdvig} for the phenomenological
$\bar{K}N - \pi \Sigma$ potentials. Keeping in mind the results for
the $K^- d$ scattering lengths discussed in the previous section it is
an expected result since the $1s$ level shift and width of an hadronic
atom are directly connected to the strong scattering length of the system.
However, the results obtained using three different models
of $\bar{K}N$ interaction are rather close to each other. We think, that
the important point here is the fact that all three potentials reproduce
low-energy experimental data on $K^- p$ scattering and kaonic hydrogen with
the same level of accuracy.

We checked the accuracy of the approximate corrected Deser
formula, allowing simple computation of characteristics
of a kaonic atom from a known scattering length. The result obtained
using the $a_{K^- d}$ value from Table~\ref{aKd.tab}:
\begin{equation}
\Delta E^{Chiral}_{K^- d, cD} = -878 \; {\rm eV,} \qquad 
 \Gamma^{Chiral}_{K^- d, cD} = 724 \; {\rm eV,}
\end{equation}
compared to more accurate ones $\Delta E^{Chiral}_{K^- d}$,
$\Gamma^{Chiral}_{K^- d}$ from the same Table show that 
in this case the error of the approximate formula is as large
as for the case of phenomenological $\bar{K}N$ potentials.
As in~\cite{my_Kd_sdvig}, the corrected Deser formula underestimates
the width of the $1s$ level of kaonic deuterium by $~30\%$. Therefore,
the validity of this statement does not depend on the model of
the $\bar{K}N$ interaction.

We would like to note that our results for the $\Delta E^{Chiral}_{K^- d}$
and $\Gamma^{Chiral}_{K^- d}$, shown in Table~\ref{aKd.tab} can not be
called ``exact'', but only ``accurate'' since the $1s$ level shift and width
were obtained from the two-body calculation with a point-like deuteron,
interacting with a kaon through the complex potential.
It means that the size of deuteron was taken into account only effectively
through the potential, which reproduces the three-body $K^- d$ AGS amplitudes.
As for the corrected Deser formula, it contains no three-body information
at all since the only input is a $K^- d$ scattering length. Moreover, the formula
relies on further approximations, which are absent in our calculation,
and give a $~10\%$ error already for the two-body case.

\section{Conclusions}
\label{conclusions.sect}

We constructed three-channel isospin dependent chirally-motivated
$\bar{K}N - \pi \Sigma - \pi \Lambda$ potential and used it in the
Faddeev-type calculations of the low-energy elastic $K^- d$ amplitudes,
including $K^- d$ scattering length, and effective range. The potential
reproduces all low-energy experimental data on $K^- p$ scattering and
characteristics of kaonic hydrogen with the same level of accuracy as
our phenomenological potentials with one- and two-pole structure of
$\Lambda(1405)$ resonance. Comparison of the results allows to reveal
the effect of the three different models of $\bar{K}N$ interaction used
in the three-body calculations. It turnes out that low-energy $K^- d$
elastic amplitudes and characteristics of kaonic deuterium obtained with
the three potentials are rather close. Therefore, comparison with
eventual results of an experiment on kaonic deuterium hardly could
choose one of the models of $\bar{K}N$ interaction. Additionally, we
found no quasi-bound states in the $K^- d$ system.

\vspace{5mm}

\noindent
{\bf Acknowledgments.}
The work was supported by the Czech GACR grant P203/12/2126 and the Hungarian
OTKA grant 109462. One of the authors (NVS) is thankful to J. Haidenbauer for
his comments on some details of $\bar{K}N$ models of interaction.


\begin{thebibliography}{99}

\bibitem{my_Kd} N.V. Shevchenko, Phys. Rev. C 85, 034001 (2012).

\bibitem{my_Kd_sdvig} N.V. Shevchenko, Nucl. Phys. A 890-891, 50 (2012).

%-------------- Kd Faddeev results:
\bibitem{Kd_TGE} G. Toker, A. Gal, J.M. Eisenberg, Nucl. Phys. A 362, 405 (1981).

\bibitem{Kd_TDD} M. Torres, R.H. Dalitz, A. Deloff, Phys. Lett. B 174, 213 (1986).

\bibitem{Kd_Deloff} A. Deloff, Phys. Rev. C 61, 024004 (2000).

\bibitem{Kd_BFMS_new} T. Mizutani, C. Fayard,  B. Saghai, K. Tsushima,
Phys. Rev. C 87, 035201 (2013).

\bibitem{SIDDHARTA-2} C. Curceanu {\it et al.}., Nucl. Phys. A 914, 251 (2013).

%---------------- Kiral'schina:
\bibitem{Borasoy_aKp} B. Borasoy, U.-G. Mei{\ss}ner, R. Ni{\ss}ler,
Phys. Rev. C {\bf 74}, 055201 (2006).

\bibitem{cieply2012} A. Ciepl\'y, J. Smejkal, Nucl. Phys. A 881, 115 (2012).

\bibitem{oller} Z.-H. Guo, J.A. Oller, Phys.Rev. C 87, 035202 (2013).

\bibitem{ihw} Y. Ikeda, T. Hyodo, W. Weise, Nucl.Phys. A 881, 98 (2012).

\bibitem{cieply2007} A. Ciepl\'y, J. Smejkal, Eur. Phys. J A 34, 237 (2007).
%----------------------------------------------------------------------

\bibitem{corDeser} U.-G. Mei{\ss}ner, U. Raha, A. Rusetsky,
Eur. Phys. J. C 35, 349 (2004).

\bibitem{ourPRC_isobreak} J. R\'evai, N.V. Shevchenko,
Phys. Rev. C 79, 035202 (2009).

\bibitem{Kp2exp} M. Sakitt {\it et al.}, Phys. Rev. 139, B719 (1965).

\bibitem{Kp3exp} J.K. Kim, Phys. Rev. Lett. 14, 29 (1965);
%Columbia University Report, Nevis, 149 (1966); 
Phys. Rev. Lett. 19, 1074 (1967).

\bibitem{Kp4exp} W. Kittel, G. Otter, and I. Wacek, Phys. Lett. 21, 349 (1966).

\bibitem{Kp5exp} J. Ciborowski {\it et al.}, J. Phys. G 8, 13 (1982).

\bibitem{Kp6exp} D. Evans {\it et al.}, J. Phys. G 9, 885 (1983).

\bibitem{Kp1exp} W.E. Humphrey, R.R. Ross, Phys. Rev. 127, 1305 (1962).

\bibitem{SIDDHARTA} M. Bazzi {\it et al.} (SIDDHARTA Collaboration),
Phys. Lett. B 704, 113 (2011).

\bibitem{gammaKp1} D.N. Tovee {\it et al.}, Nucl. Phys. B 33, 493 (1971).

\bibitem{gammaKp2} R.J.~Nowak {\it et al.}, Nucl. Phys. B 139, 61 (1978).

\bibitem{Janos} J. R\'evai, Few-Body Syst. 54, 1865 (2013).

\bibitem{log_sing} F. Sohre and H. Ziegelman, Phys. Lett. B 34, 579 (1971).

\bibitem{KEK} T.M.~Ito {\it et al.}, Phys. Rev. C 58, 2366 (1998).

\end{thebibliography}
\end{document}